\documentclass [12pt]{article}
\usepackage{lscape}                                           %
\usepackage[centertags]{amsmath}
\usepackage{amsfonts} \usepackage{amssymb}\usepackage{eufrak}
\usepackage{colordvi}
\usepackage{epsfig}
\usepackage{a4}
\usepackage{cite}
\usepackage{lscape}
\newcommand{\COMMENTO}[1]{}

\begin{document}
\begin{titlepage}
%\rightline{DSF-25-2008}
\rightline{NORDITA-2008-49} \vskip 3.0cm
\centerline{\LARGE \bf  Discussing string extensions of the 
}\vskip .5cm
\centerline{\LARGE \bf Standard Model  in  D brane world}
%\vskip .5cm\centerline{\LARGE \bf in String Theory }
\vskip 1.0cm \centerline{\bf P. Di Vecchia$^{a,b}$
%, A. Liccardo$^{c,d}$, R. Marotta$^d$
%I. Pesando $^d $
%and F. Pezzella$^d$
}
\vskip .6cm \centerline{\sl $^a$
The Niels Bohr Institute, Blegdamsvej 17, DK-2100 Copenhagen \O, Denmark}
\vskip .4cm
\centerline{\sl $^b$
Nordita, Roslagstullsbacken 23, SE-10691 Stockholm, Sweden}
%\vskip .4cm \centerline{\sl
 %$^c$ Dipartimento di
%Scienze Fisiche, Universit\`a ``Federico II'' di Napoli} 
%\centerline{\sl Complesso Universitario Monte
%S. Angelo ed. 6, via Cintia,  I-80126 Napoli, Italy}
%\vskip .4cm
%\centerline{\sl 
%$^d$ Istituto Nazionale di Fisica Nucleare, Sezione di Napoli}
%\centerline{ \sl Complesso Universitario Monte
%S. Angelo ed. 6, via Cintia,  I-80126 Napoli, Italy}

%\vskip 0.4cm \centerline{ \sl $^d$ Dipartimento di Fisica Teorica,
%  Universit\`a di Torino and INFN, Sezione di Torino,} \centerline{\sl
% via P. Giuria 1,  I-10125, Torino, Italy}
 \vskip 1cm

\begin{abstract}
In this talk we will describe the problems that one encounters when one tries to 
connect string theory with particle phenomenology.  Then, in order to have chiral 
matter describing quarks and leptons, we introduce the magnetized D branes.
Finally, as an  explicit example of a string extension of the Standard Model, we
 will
describe the one constructed by Ib{\'{a}}{\~{n}}ez, Marchesano and Rabad{\'{a}}n.
\end{abstract}
\vfill  {\small{ Work partially supported by the European
Community's Human Potential Programme under contract
MRTN-CT-2004-005104 ``Constituents, Fundamental Forces and Symmetries
of the Universe''. }}
\end{titlepage}

\newpage

%%%%%%%%%%%%%%%%%%%%%%
\tableofcontents       %
\vskip 1cm             %
%%%%%%%%%%%%%%%%%%%%%%%

\section{String theory and experiments}
\label{stringexpe}

The strongest motivation for string theory is the fact that it provides
a consistent quantum theory of gravity unified with  
gauge interactions. This is a consequence of the fact that  
string theory has a  parameter  $\alpha'$, related to the 
string tension by $T = \frac{1}{2 \pi \alpha'}$, 
of  dimension of a $(length)^2$ 
that acts as an ultraviolet cutoff  $\Lambda = \frac{1}{\sqrt{\alpha'}}$.
Because of this physical cutoff  all loop integrals are finite in the UV.

The presence of this dimensional 
parameter  $\alpha'$  implies that string theory can be viewed  as 
an extension, rather than an alternative,  to  field theory 
pretty much as special relativity and quantum mechanics
are an extension of respectively, Galilean  and Classical mechanics. 
Those latter theories can be  obtained from the former ones by taking
respectively  the limit   $c \rightarrow \infty$ and  
$h \rightarrow 0$. Analogously , by taking the zero slope 
limit ($\alpha' \rightarrow 0$), one can see that perturbative 
string theory reduces to
a perturbative  gauge field theory unified with an extension of 
Einstein's theory of general
relativity and that, in this limit, one recovers   the UV divergences 
of perturbative quantum gravity unified with gauge theories. 
As it is known since long time,  they are due
to the point-like structure of the elementary constituents~\cite{landau}.  

But, if string theory has something to do with Nature, how can we see 
stringy effects
in experiments? The answer to this question depends of course on the
energy $E$ available. If  $\alpha' E^2 <<1$, then one will see only the 
limiting field theory. In other words, if we knew  the strength of 
$\alpha'$ we could tell  at what energy  one would see stringy  effects
that will manifest as deviations  from the  field theoretical behaviour
 based on  point-like objects. If $\frac{1}{\sqrt{\alpha'}} \sim 10\,  TeV$, then
 stringy effects can be seen in present experiments, while, if 
 $\frac{1}{\sqrt{\alpha'}} >> 10 TeV$, then the presence of a string theory
 cannot be directly seen.  
 
Having said this, let us now discuss where we stand in string theory.  
The simplest string theory is the bosonic string that is, however, not 
immediately consistent because it contains tachyons in the spectrum.
Around 1985 it was clear that we have  5 ten-dimensional  consistent 
string theories: $IIA, IIB, I,$ Heterotic $E_8 \times E_8$ and Heterotic
 $SO(32)$.
They are  {inequivalent} in string perturbation theory 
($g_s < 1$), {supersymmetric} and  {unify
in a consistent quantum theory gauge theories with
gravity}. It must be said, however, that,  unlike $\alpha'$,  the string 
coupling constant $g_s$ {is not} 
a parameter to be fixed from experiments. In fact,
it corresponds to the vacuum expectation value of a string  excitation,
called the dilaton, {$g_s = {\rm e}^{\langle \phi \rangle}$}, 
that  {should be  fixed} by the
minima of the  {dilaton potential}.
But the potential for the dilaton is {flat} in any order of string 
perturbation theory and therefore, for each value of $\langle \phi \rangle$, 
we have an inequivalent theory.
This is  clearly  unsatisfactory for a theory, as string theory, that 
has the potential   to explain  everything. This is, however,  not the only problem!
In fact,  if string theory is the fundamental theory unifying all 
interactions, why do we have 5 theories instead of just one?
The key to solve this problem came from the discovery  in string theory
of  new $p$-dimensional states, called  {D(irichlet) p branes~\cite{9510017}. 
In the following we will explain their origin.

The spectrum of massless states of the II
    theories is given in the table
\begin{center}
\begin{tabular}{|lllr|}
\hline $G_{\mu\nu}$ & $B_{\mu\nu}$& $\phi$ & {\small NS-NS sector}\\
{\small Metric}   & {\small Kalb-Ramond} 
& {\small Dilaton}&\\
$C_0,\,C_2$& $C_4,\,C_6$& $C_8$& {\small RR sector\,\, IIB}\\
$C_1,\,C_3$& $C_5$ & $C_7$ &  {\small RR sector\,\, IIA} \\
\hline
\end{tabular}
\end{center}
where the RR $C_i$ stands for an antisymmetric tensor potential  
 $C_{\mu_1 \mu_2 \dots \mu_i}$ with $i$ indices.

These antisymmetric potentials  are generalizations of the electromagnetic
    potential $A_{\mu}$ \\
\begin{eqnarray}
\int A_{\mu} dx^{\mu} \Longrightarrow 
\int A_{\mu_1 \mu_2 \dots \mu_{p+1} } 
 \frac{ 
d x^{\mu_1}\wedge  d x^{\mu_2}  \dots \wedge  dx^{\mu_{p+1} }  }{ (p+1)! }
\label{Adx}
\end{eqnarray}
In fact, as the electromagnetic field is coupled to point-like particles, 
so they are coupled to p-dimensional objects.  

There exist classical solutions of the
    low-energy string effective action that are coupled to the metric, the
    dilaton and are charged with respect a RR field. The RR potential behaves
    as 
\begin{center}
{$C_{01 \dots p} \sim \frac{1}{r^{d-3-p}} \Longleftrightarrow C_0 \sim
 \frac{1}{r} \,\,$ if $d=4, p=0$}
\end{center}
that generalizes to a p-dimensional object in a d-dimensional space-time 
the behaviour
of the Coulomb potential valid for a point-particle in four dimensions. 
They are additional non-perturbative states of string theory with
 tension and RR charge given by:
\begin{eqnarray}
\tau_p  &=& \frac{Mass}{p-volume}= \frac{(2\pi \sqrt{\alpha'})^{1-p}}{2
  \pi \alpha' g_s } \\ \nonumber
 \mu_p &=& \sqrt{2 \pi} (2\pi \sqrt{\alpha'})^{3-p}
\label{tenscha}
\end{eqnarray} 
Being  non-perturbative objects  their tension diverges in the 
perturbative limit $(g_s \rightarrow 0)$.

They are called {D(irichlet) p branes} because they have
 open strings attached to their (p+1)-dim. world-volume:
\begin{eqnarray}
 \partial_{\sigma} X^{\mu} ( \sigma =0, \pi ; \tau  )
&=&0~~\mu=0 \dots p\nonumber \\
\partial_{\tau} X^{i} (\sigma=0, \pi;  \tau ) &=& 0~~i =p+1  \dots 10
\label{dirineu}
\end{eqnarray}
As follows from the previous equations, the open string  satisfies 
Neumann boundary conditions along the world-volume of the 
Dp brane and Dirichlet boundary conditions in the directions transverse 
to the world-volume of the Dp brane.
Remember that the motion of a string is described by the string coordinate 
{$X^{\mu} (\sigma, \tau)$} that is a function of the parameters $\sigma$ and $\tau$
that parameterize  the  world-sheet of a string. This parameterization is such that 
 {$\sigma =0, \pi$} correspond 
to the  {two end-points} of an open string. It turns out  that, in a brane world,
the states corresponding to the excitations of open strings  
live in the (p+1)-dim. world-volume of a Dp brane, while those 
corresponding to the
excitations of  closed strings 
 live in the entire ten dimensional space. This  means that the gauge
 theories, described by open strings, live on the world-volume of a Dp brane, while gravity, described by closed strings, lives in 
 the entire ten-dimensional space-time. In particular,
if we have a stack of $N$ parallel D branes, then we have 
$N^2$ open strings having their endpoints on the D branes: these are the
degrees of freedom of the adjoint representation of $U(N)$. One concludes 
that the  open strings attached to the same stack of D branes transform according 
to the adjoint representation of $U(N)$. The massless string excitations correspond 
to the  gauge fields of $U(N)$. Therefore a stack of N D branes has a   
{$U(N) = SU(N) \times U(1)$  
gauge theory} living on their world-volume.

The discovery of Dp branes opened the way in 1995 to the discovery of 
the string dualities and this led to understand that the 
5 string theories were actually part of a unique  11-dimensional theory,  
called {M theory}.

However, in the experiments we observe only  {4} and not 
10 or 11 non-compact
directions. Therefore 6 of the 10 dimensions must be compactified and small: 
{$ R^{1,9}  \rightarrow R^{1,3} \times M_6 $}, where
{ $M_6$ is a compact manifold}.
In order to preserve at least $N=1$ supersymmetry $M_6$ must 
be a Calabi-Yau manifold.
But this means that the low-energy physics will depend not only on 
{$\alpha'$} and  {$g_s$} , but also on the size and the shape 
of the manifold $M_6$.

Originally the most promising string theory for phenomenology was considered
the Heterotic $E_8 \times E_8$ that was studied intensively.
But in this theory both the fundamental string length $\sqrt{\alpha'}$ and
the size of the extra dimensions are  of the order of the Planck
length:
\begin{eqnarray}
&&\frac{1}{\sqrt{\alpha'}}  \equiv M_s 
= \frac{ M_{Pl.} \sqrt{\alpha_{GUT} } }{2} \sim
\frac{ M_{Pl.}  }{10 }  \nonumber \\ 
&&\frac{R}{\sqrt{\alpha'}} \sim 1~~;~~ if\,\,  g_s < 1 
\label{hete}
\end{eqnarray}
They are both too  small to be directly observed in present and even 
future experiments!  This means that, if we want to compare 
perturbative heterotic string theory with experiments, we need to have 
a very good control of the theory to be able to extrapolate
to low energy.

Later on in 1998  it became clear that in type I and in a brane world
one could allow for {much larger values} for the string 
length  {$\sqrt{\alpha'}$}
and for the {size} of  the extra dimensions  without being in contradiction
with the experimental data~\cite{98}. However, it is not clear if Nature likes 
these larger values.

When we compactify 6 of the 10 dimensions, in addition to the dilaton,
we generate a bunch of scalar fields {(moduli)}  
corresponding  to the components
of the metric and of the other closed string fields in the extra dimensions.}
{Their vacuum expectation 
values, { corresponding to the parameters of the compact manifold}, 
are not fixed
in any order of perturbation theory  {because their potential is flat}.
{We get a continuum of string vacua for any value of the moduli and   
this is obviously   not  good for phenomenology. In order to compare
string theory with particle phenomenology one needs to find mechanisms
to stabilize the moduli.
In the last few years a lot of progress has been made 
in this direction because
one has been able to stabilize them  by 
the introduction of non-zero fluxes for some of the NS-NS and R-R fields.

But we are still left with a    {discrete} (and  {huge}) 
quantity of string vacua and this problem goes under the name of 
the  { "Landscape Problem".}

The question is now:{ how do we fix the vacuum we live in?}
Do we need to rely on the  Anthropic principle or maybe can  this problem
be solved by reaching a better understanding of string theory?

Rather than to discuss these two alternatives it is, in my opinion, more
useful to try to construct string extensions of the Standard model (SM) and of 
the Minimal Supersymmetric Standard Model (MSSM). This has been called a 
bottom-up approach because one does not derive the SM or the MSSM 
from string theory as one would do in a more ambitious 
top-down approach, but instead, 
since we know that the SM correctly describes Nature
at   the energy reached up to now, one tries to see if the SM 
can be consistently incorporated in string theory.

If we want to construct string extensions of the SM 
 in an explicit way we must limit ourselves 
to toroidal  compactifications with orbifolds and orientifolds
and, most important, we 
need to have massless  open strings 
corresponding to   chiral fermions in four dimensions for describing 
quarks and leptons.

The simplest explicitly solvable models with chiral matter in four dimensions 
  are those based on several stacks of 
{intersecting branes} or of their T-dual  {magnetized branes}  
on {$R^{3,1} \times {{T}}^2 \times {{T}}^2 \times {{T}}^2$}.
These are the models that we are going to describe in the following section.

\section{Magnetized D branes}
\label{magne}

Magnetized branes are characterized by having a non-zero
constant magnetic field  along the six compact directions of the torus 
$T^2 \times T^2 \times T^2$.
Let us assume that we have two stacks of  magnetized D branes  that we call 
stack a and stack b.  We want to study the motion and the spectrum of
the open strings attached with one end-point to the stack a and with the 
other end-point to stack b when the magnetizations on the two 
stacks are different from each other. 
This kind of open strings are called twisted, dycharged or chiral
strings. Their motion is described by the following action:
\begin{eqnarray}
S = S_{bulk} + S_{boundary}
\label{spluss}
%\nonumber
\end{eqnarray}
where
\begin{eqnarray}
S_{bulk} = - \frac{1}{4 \pi
  \alpha'}
\int d\tau \int_{0}^{\pi} d \sigma
\left[ G_{ab} \partial_{\alpha} X^a \partial_{\beta} X^b \eta^{\alpha \beta}
     - B_{ab} \epsilon^{\alpha \beta} \partial_{\alpha} X^a
\partial_{\beta} X^b \right] 
%\nonumber
\label{acti853}
\end{eqnarray}
and
%being the world-sheet metric $\eta_{\alpha \beta} =
%\mbox{diag}(-1,1)$ and
%$\epsilon^{01} =1$, while   $S_{boundary}$ is:
\begin{eqnarray}
&&S_{boundary}   
 =- q_a \int d \tau A_{i}^{(a)} \partial_{\tau} X^{i}
|_{\sigma =0} +
q_{b} \int d \tau A_{i}^{(b)} \partial_{\tau} X^{i}
|_{\sigma =\pi} = \nonumber \\
&&=
 \frac{q_a}{2} \int d \tau F_{i j}^{(a)} X^j
  {\dot{X}}^{i}|_{\sigma =0} 
 -\frac{q_{b}}{2} \int d \tau F_{i j}^{(b)} X^j
  {\dot{X}}^{i}|_{\sigma=\pi}
\label{s2}
%\nonumber
\end{eqnarray}
The two gauge field strengths $F_{i j}^{(a, b)} $ are constant and we choose 
the gauge  in which the vector potentials are given by:
\begin{eqnarray}
A_{i}^{(a, b)} =-\frac{1}{2} F_{i j}^{(a, b)} x^j~.
\label{A-F-main}
%\nonumber
\end{eqnarray}
The data of the torus ${{T}}^2$, {called moduli}, are included 
in the constant   $G_{ij}$, that is  the metric of the torus $T^2$,  and 
$B_{ij}$}, that is a background two-index antisymmetric Kalb-Ramond field. 
They are related to the   {complex and K{\"{a}}hler
structures} of the torus $T^2$:
\begin{eqnarray}
U &\equiv&  U_1 + i U_2 = \frac{G_{12}}{G_{11}} +i \frac{\sqrt{G}}{G_{11}}
\nonumber \\
T &\equiv& T_1 + i T_2 = B_{12} + i \sqrt{G}
%\nonumber
\label{UT}
\end{eqnarray}
given by
\begin{eqnarray}
\label{GB}
G_{ij}
&=&  \frac{T_2 }{U_2}\,
\left(\begin{array}{cc} 1  & U_1  \\
                 U_1  & |U |^2\end{array} \right) 
                 %\mbox{;}
\nonumber \\
B_{ij}
&=& \left( \begin{array}{cc}
0 & - T_1  \\ 
T_1  & 0 
\end{array}  \right)
%\nonumber
\end{eqnarray}
They are the closed string moduli~\footnote{In this talk, for the sake of simplicity,
we take $T_1=0$}.
 
F is  constrained by the fact that its  flux  
 is an integer: 
%(\structure{$0 \leq x^{1,2} \leq 2 \pi \sqrt{\alpha'}$)}
\begin{eqnarray}
\int  Tr \left(\frac{q F}{2 \pi} \right) = m  \Longrightarrow 
2 \pi \alpha' q   F_{12}    = \frac{m}{n}
%~~~;~~~F \equiv F_{ij} \frac{dx^i \wedge dx^j}{2}
%\nonumber
\label{F}
\end{eqnarray}
They are the open string moduli.
The D brane { is wrapped  $n$} times on the torus and 
the flux of $F$, on a compact space as $T^2$, must be  {
an integer $m$ corresponding to a
magnetic charge.
%The fact that the brane \alert{is wrapped n times on the torus} 
%is described by introducing 
%a non-trivial $U(n)$ gauge bundle.

The most general motion of an open string in this constant  
background  can be explicitly determined and the theory can 
be explicitly quantized~\cite{BP9209}. 

One gets a string extension of the motion of an electron in a constant 
magnetic field on a torus. Also in the string case
the ground state is degenerate and the degeneracy is given 
by the number of Landau levels.
One can also see that, when $\alpha' \rightarrow 0$, one goes back 
 to the problem of an electron in a constant magnetic field.
 
The mass spectrum of the string states can be exactly determined
and it is given by:
\begin{eqnarray}
\alpha' M^2 = N_{4}^{X} + N_{4}^{\psi} + 
N_{comp.}^{X} + N_{comp}^{\psi} + 
 \frac{x}{2} \sum_{i=1}^{3} \nu_i - \frac{x}{2}
%\nonumber
\label{mass2}
\end{eqnarray}
where
{$x =0$  for fermions} (R sector)  and  {$x = {1}$ for bosons}
  (NS sector) and  
\begin{eqnarray}
N^{X}_{4} = \sum_{n=1}^{\infty} n a_{n}^{\dagger} 
\cdot a_{n}~~;~~
N^{\psi}_{4} = \sum_{n=\frac{x}{2}}^{\infty} 
n  \psi_{n}^{\dagger} \cdot \psi_{n}
%\nonumber
\label{NN}
\end{eqnarray}
\begin{eqnarray}
N^{X}_{comp} = \sum_{r=1}^{3} \left[
\sum_{n=0}^{\infty} (n+ \nu_r) 
a_{n+ \nu_r}^{\dagger (r)}  a_{n+\nu_r}^{(r)} +
 \sum_{n=1}^{\infty} (n- \nu_r) 
{\bar{a}}_{n- \nu_r}^{\dagger (r)}  {\bar{a}}_{n- \nu_r}^{(r)}  \right]
%\nonumber
\label{NX}
\end{eqnarray}
\begin{eqnarray}
N^{\psi}_{comp} = \sum_{r=1}^{3} \left[
\sum_{n=\frac{x}{2}}^{\infty} (n+ \nu_r) \psi_{n+ \nu_r}^{\dagger (r)}  
\psi_{n+\nu_r}^{(r)}  
+ \sum_{n=1 - \frac{x}{2}}^{\infty} (n- \nu_r) 
{\bar{\psi}}_{n- \nu_r}^{\dagger (r)}  {\bar{\psi}}_{n- \nu_r}^{(r)}  \right]
%\nonumber
\label{Npsi}
\end{eqnarray}
where
\begin{eqnarray}
{\nu_r = \nu_{r}^{(a)} - \nu_{r}^{(b)}}~~~;~~~
{\tan \pi \nu_{r}^{a, b}  
%2 \pi \alpha' q_{0, \pi} \frac{F_{12}^{(0,\pi)i}}{T_{2}^{(i)}}
= \frac{m_{r}^{(a, b)}}{n_{r}^{(a, b)} T_{2}^{(r)}}}
%\nonumber
\label{nus}
\end{eqnarray}
$T_{2}^{(r)}$ is the volume of the $r$-th  torus.

In the fermionic  sector the lowest state is the vacuum state.
It is a {4-dimensional massless chiral spinor}. This is due to the fact that the ten-dimensional GSO projection reduces to the four-dimensional one 
because the fermionic zero modes are absent in the six-dimensional 
compact directions.

For generic values of $\nu_1 , \nu_2, \nu_3$
 there is no massless state in the bosonic  sector and this means that, in general, 
 the original 10-dim supersymmetry is broken~\cite{9503030}.

The lowest bosonic states are  
\begin{eqnarray}
 {\bar{\psi}}_{\frac{1}{2} - \nu_r}^{\dagger (r)} | 0 >~~;~~r=1,2,3
 \label{r}
 \end{eqnarray}
 with  masses, respectively given by
 \begin{eqnarray} 
 \alpha' M^2 = \frac{1}{2} \sum_{s=1}^{3}  \nu_s - \nu_r~~;~~r=1,2,3 
 \label{mr}
 \end{eqnarray}
 and 
\begin{eqnarray}
{\bar{\psi}}_{\frac{1}{2} - \nu_1 }^{\dagger (1) }  
{\bar{\psi}}_{\frac{1}{2} - \nu_2 }^{\dagger (2) }
{\bar{\psi}}_{\frac{1}{2} - \nu_3 }^{\dagger (3) }  | 0 >
\label{mrb}
\end{eqnarray}
with mass given by
\begin{eqnarray} 
  \alpha' M^2 = \frac{2 - \nu_1 - \nu_2 - \nu_3}{2}
\label{lowbos}
\end{eqnarray}
%For $i=1$ its mass is vanishing if \alert{$\nu_1 = \nu_2 + \nu_3$}.  
One of these states becomes massless if one of the following 
identities is satisfied:
\begin{eqnarray}
&&\nu_1 = \nu_2 + \nu_3~~;~~\nu_2 = \nu_1 + \nu_3
\nonumber \\
&&\nu_3 = \nu_1 + \nu_2
~~;~~\nu_1 + \nu_2 + \nu_3 =2
%\nonumber
\label{susy8}
\end{eqnarray}
In each of these  cases a  four-dimensional 
{${\cal{N}}=1$ supersymmetry 
is restaured!}

{In general the ground state for the open strings, having their end-points, 
respectively on  stacks a and b,
 is degenerate.}

Its degeneracy is given by the number of Landau levels
as in the case of
a point-like particle:
\begin{eqnarray}
I_{ab}  = - \prod_{i=1}^{3} \left\{ n_{i}^{(a)} n_{i}^{(b)}   
\int \left[ \frac{q_a F^{(a)}_{i}  - q_{b} F^{(b)}_{i}}{2 \pi}\right] \right\}
%=\nonumber \\
= \prod_{i=1}^{3} \left[ m_{i}^{(b)}  n_{i}^{(a)}  -m_{i}^{(a)}  n_{i}^{(b)}  
\right]  
%\nonumber
\label{Landau}
\end{eqnarray}
that gives the  {number of families} in the 
phenomenological applications. Note that $I_{ab}$ can be both positive 
and negative. The convention is that a positive $I_{ab}$ describes left-handed 
fermions, while a negative  $I_{ab}$ describes right-handed fermions.

In this section we have considered D9 branes on $T^2 \times T^2 \times T^2$
with three magnetizations $\nu_1 , \nu_2 , \nu_3$. This system is T-dual 
to a system of non-magnetized, but intersecting  D6 branes wrapped on 
the   three one-cycles $[a_r ]$ of the three tori with wrapping numbers $n_{r}$.
In this T-dual picture the magnetizations become the angles in the three tori
between the D6 branes and the number of Landau levels become
the  number of intersections. In the following, we will mostly use the language
of the D6 branes instead of that of the magnetized branes.
 
\section{An example of  string extension of the Standard Model}
\label{example}
 
In this section we briefly describe a consistent string extension of 
the SM constructed in Ref.~\cite{0105155}. In order to have a
string extension of the  SM we need to introduce four 
different stacks
of magnetized D9 branes that we call: $a, b, c, d$. 
The stack $a$ consists of three branes, called the baryonic branes,
that have on their world-volume a $U(3)= SU(3) \times U(1)$ 
gauge theory, where the $SU (3)$ subgroup is the color $SU (3)$ 
describing strong interactions. The stack $b$ consists of two branes, called
the left branes, that have on their world-volume a $U(2) = SU(2) \times U(1)$
gauge theory with the gauge group $SU(2)$  being the 
$SU(2)_L$ of the SM. The stacks $c$ and $d$ consist of 
one brane each, called respectively, the leptonic and the right branes. They 
have both a $U(1)$ theory living on their world-volume. In conclusion,
on this system of branes we have the following gauge group:
\begin{eqnarray}
 SU(3)_{a}  \times SU(2)_{b}   \times U(1)_{a}  \times U(1)_{b} 
\times U(1)_{c} \times U(1)_{d}
\label{gg}
\end{eqnarray}
where $SU(3)_a$ is the color $SU(3)$ and $SU(2)_b$ is the $SU(2)_L$.
In addition to these two groups of the SM we have four $U(1)$ instead of the
just the $U(1)$ hypercharge.  In order to determine the $U(1)$ hypercharge and
to see the role of the other $U(1)$'s we need first to discuss the cancellation
of the gauge anomalies. In fact, in a model, like the one discussed above, 
containing
chiral fermions, we need to check that all the gauge anomalies are cancelled.
This cannot be realized in the model just discussed above, but we have to add
to it an orientifold projection. This means that for each D brane we need to 
introduce its image that we denote  with a star. Therefore for each of the
four stacks of  D branes $a,b,c,d$ we will have their images denoted with
$a^* , b^* , c^* , d^* $. 

In our orientifold theory with intersecting D6 branes the tadpole cancellation
condition reads~\footnote{Remember that in this talk we take $T_1 =0$. }:
\begin{eqnarray}
\sum_{A=1}^{K} N_A \left( \Pi_A + \Pi_{A^*} \right) = 32 \Pi_{O6}  
\label{ori}
\end{eqnarray}
In the case of the torus $T^2   \times T^2   \times  T^2$ the three-cycles
$\Pi_A , \Pi_{A^*},  \Pi_{O6}$ are given by:
\begin{eqnarray}
&&\Pi_A = \prod_{r=1}^{3} \left[ n_{A}^{(r)} [ a_r ] +  
m_{a}^{(r)} [ b_r ]\right] \nonumber \\
&&\Pi_{A^*}  = \prod_{r=1}^{3} \left[ n_{A}^{(r)} [ a_r ] -  
m_{A}^{(r)} [ b_r ]\right] \nonumber\\
&&\Pi_{O6} =  \prod_{r=1}^{3} [a_r]
\label{pia}
\end{eqnarray}
where $[a_r]$ and $[b_r ]$ are the two cycles of the r-th torus
satisfying the following relations:
\begin{eqnarray}
&&[a_r ] \cdot [a_s ] = [b_r ] \cdot [b_s ] =0 \nonumber \\
&&[a_r ] \cdot [b_s ] = -
[b_r ] \cdot [a_s ] = \delta_{rs}
\label{inte45}
\end{eqnarray}
Using the previous equations it is possible to compute the following quantities:
\begin{eqnarray}
&& \Pi_A  \cdot \Pi_B = I_{AB} \equiv \prod_{r=1}^{3} 
\left[n_{A}^{(r)} m_{B}^{(r)} -
n_{B}^{(r)} m_{A}^{(r)} \right] \nonumber \\
&& \Pi_A \cdot \Pi_{O6}  = -
 \prod_{r=1}^{3} m_{A}^{(r)}=  I_{A O6} 
\label{pipi}
\end{eqnarray}
and
\begin{eqnarray}
\Pi_A \cdot \Pi_{B^*} =  - \Pi_{A^*} \cdot \Pi_B  \equiv I_{AB^*} = 
 -   \prod_{r=1}^{3} \left[n_{A}^{(r)} m_{B}^{(r)} +
n_{B}^{(r)} m_{A}^{(r)} \right] 
\label{piab}
\end{eqnarray}
Before we proceed, let us summarize the spectrum of open strings stretched
between intersecting D6 branes~\cite{0007024}. 
The open strings attached with one end-point at the stack A and 
with the other end-point at the stack B  transform according to the 
bifundamental representation $(N_A , {\bar{N}}_B )$ of the gauge 
groups $U(N_A)$ and $U(N_B )$ respectively, and their number is equal
to $I_{AB} = \Pi_A \cdot \Pi_B$:
\begin{eqnarray}
&&(A,B) \rightarrow (N_A , {\bar{N}}_B )  \nonumber \\
&&I_{AB}=  \prod_{r=1}^{3} \left[n_{A}^{(r)} m_{B}^{(r)} -
n_{B}^{(r)} m_{A}^{(r)} \right]
\label{ab}
\end{eqnarray}
The massless chiral fermions 
corresponding to open strings stretched between a 
stack $A$ and the image $B^*$  of the stack b transform according 
to the bifundamental representation $(N_A , N_B )$ of the gauge groups 
$U(N_A )$ and $U(N_B )$ and their number
is equal to $I_{AB^*}$:
\begin{eqnarray}
&& (A, B^*) \rightarrow (N_A , N_B ) \nonumber \\
&&I_{A B^*} = 
-\prod_{r=1}^{3} \left[n_{A}^{(r)} m_{B}^{(r)} +
n_{B}^{(r)} m_{A}^{(r)} \right]
\label{ab'}
\end{eqnarray}
Finally the massless chiral fermions corresponding to open strings having
one end-point attached to the  stack $A$ and and the other one to 
its image $A^*$  transform according to both 
the two-index symmetric and two-index antisymmetric representation
of the gauge group $U(N_A )$.  Their total number is given by 
\begin{eqnarray}
I_{A A^*} = -8 \prod_{r=1}^{3} n_{A}^{(r)} m_{A}^{(r)}
\label{aa'}
\end{eqnarray}
Their multiplicity is given respectively by:
\begin{eqnarray}
&& A = - 4  \prod_{r=1}^{3}  m_{A}^{(r)} 
\left( \prod_{r=1}^{3} n_{A}^{(r)} +1  \right) \nonumber \\
&& S = - 4  \prod_{r=1}^{3}  m_{A}^{(r)} 
\left( \prod_{r=1}^{3} n_{A}^{(r)} - 1  \right)
\label{AS}
\end{eqnarray}
They can also written as:
\begin{eqnarray}
&&A = \frac{1}{2} \left(I_{AA^*} + N_{O6} I_{AO6}  \right) \nonumber \\
&&S = \frac{1}{2} \left(I_{AA^*} - N_{O6} I_{AO6}  \right)
\label{ASbis}
\end{eqnarray}
Notice that the $(A,S)_A$ representations are only coupled to a single
gauge group $ U(N_A )$, while the bi-fundamentals are coupled to two
gauge groups.

Having discussed  the spectrum of chiral fermions we can now compute 
the coefficient of the non-abelian anomaly. We follow the notation of 
Ref.~\cite{0307262}
where more details can be found. The non-abelian anomaly is given by:
\begin{eqnarray}
&&{\cal{A}}_{(SU(N_A); SU(N_B )} \equiv  \sum_{r} A(r) 
= \sum_{B \neq A} N_b ( I_{AB} + I_{AB^* } ) 
+ \nonumber \\ 
&&+ \frac{1}{2} \left( I_{AA^* } + N_{O6} I_{A O6}  \right)(N_A -4) +
\nonumber \\
&&+\frac{1}{2} \left( I_{AA^*} - N_{O6} I_{A O6}  \right)(N_A + 4)
\label{noabe}
\end{eqnarray}
where the three contributions come respectively, from the fermions in
the fundamental, in the two-index antisymmetric and in the two-index
symmetric representations of the gauge group $SU(N_A )$.  $A(r)$ is 
related to  the
cubic  Casimir and is given in Table (4.1) of Ref.~\cite{0307262}.
The previous expression can be written as follows: 
\begin{eqnarray}
&&{\cal{A}}_{(SU(N_A); SU(N_B )} = \nonumber \\
&&=\sum_{B } N_B ( I_{AB} + I_{AB^*} ) -4  N_{O6} I_{A O6} = \Pi_A \times
\nonumber\\
&&  \times   
\left[ \sum_{B} N_B \left( \Pi_B + \Pi_{B^*}  \right) - 
32 \Pi_{O6} \right] =0
\label{nonabe1}
\end{eqnarray}
where $N_{O6} =8$. In the last step of the previous equation   
we have used Eq.s (\ref{pipi}) and ({\ref{piab}) and 
we have imposed the tadpole cancellation condition in Eq. (\ref{ori}). 
In conclusion, if we impose the tadpole cancellation condition we   
have automatically eliminated the non-abelian anomalies.  

The coefficient of the mixed anomalies is given by:
\begin{eqnarray}
&& {\cal{A}}_{U(1)_A; SU(N_B ) } \equiv \sum_r  Q_{A} (r) C_{B} (r)
=  \frac{1}{2} \delta_{AB} \times
\nonumber \\
&& \times \left( 
\sum_C (I_{AC} + I_{AC^* } ) N_c - Q_{O6} N_{O6} I_{A O6} \right) +
\nonumber \\
&&
 + \frac{1}{2} N_ A  (I_{AB} + I_{AB^*} )
\label{mia1}
\end{eqnarray}
where the $U(1)$ charge $Q_{A} (r)$ and the quadratic Casimir $C_{B} (r)$
are given again in Table (4.1) of Ref.~\cite{0307262}. 
If the tadpole cancellation condition is satisfied, then  the first term in Eq. 
(\ref{mia1}) is vanishing. We are left with the
second term  that we will discuss later on.
 This means that we have only fundamental representations
with degeneracy equal to $I_{AB}$.

Finally the $U(1)$ anomalies are given by:
\begin{eqnarray}
&& {\cal{A}}_{ U (1)_A; U(1)_{B}^{2}} \equiv \sum_r Q_{A} (r) Q_{B}^{2} (r) =
\frac{1}{3} \delta_{AB} N_A  \times \nonumber \\
&& \times
\left( \sum_C ( I_{AC} + I_{AC^* } ) N_C - Q_{O6} N_{O6} I_{A O6} \right)
\nonumber\\
&& + \frac{1}{2} N_A N_B ( I_{AB} + I_{AB^*})
\label{u1c}
\end{eqnarray}
The first term in the previous equations is  vanishing if we impose
the tadpole cancellation equation. We are left with the second term
that we will analyze later. 

Let us now go back to our model with four stacks of D6 brane introduced 
at the beginning of this section.
Each of the stacks will be wrapping some cycles $[\Pi_{A} ]$ 
($A= a,b,c, d$) of the product of three tori. They will intersect each 
other $I_{A  B}$  number of times. The chiral fermions live at each 
intersection and we can choose the various intersecting numbers in such a way
to reproduce the spectrum of the Standard Model. We have seen that the 
open strings    stretching between each brane and its image give rise to chiral
fermions transforming according to the double symmetric and double antisymmetric 
representation. Those states do not appear in the Standard Model and therefore
we have to impose that:
\begin{eqnarray}
I_{a a^*} = I_{b b^*} = I_{c c^*} = I_{d d^*} =0   
\label{Ialal*}
\end{eqnarray}
Since we have:
\begin{eqnarray}
I_{A  A^*} = - 8 \prod_{r=1}^{3} 
\left[ m_{A}^{(r)}  n_{A }^{(r)} \right] =0 
\label{Ialal*b}
\end{eqnarray}
This condition can be satisfied if we choose:
\begin{eqnarray}
\prod_{r=1}^{3}   m_{A}^{(r)}   =0~~;~~~A= a,b,c,d 
\label{zer9}
\end{eqnarray}
It  implies also that
\begin{eqnarray}
I_{A O6} =0
\label{aO6}
\end{eqnarray}
as follows from Eq. (\ref{pipi}). With the  choice in Eq. (\ref{zer9}) we have imposed
that both $A$ and $S$ in Eq. (\ref{ASbis}) are zero. In this way we have imposed
that there is no chiral fermion transforming according 
to the double symmetric and the double antisymmetric representations, as 
it is the case in the SM.

We have to choose the other intersecting numbers in such a way to get 
the correct spectrum of the Standard Model with three families.  
This can be realized by imposing the following intersecting numbers:
\begin{eqnarray}
I_{ab} =1 ~~~&;&~~ I_{a b^*}= 2 \\
\nonumber
I_{ac} =-3 ~~~&;&~~I_{a c^*} =-3 \\
\nonumber
I_{bd} =-3 ~~~&;&~~ I_{b d^*}= 0 \\
\nonumber
I_{cd} =3 ~~~&;&~~ I_{c d^*}= -3
\label{Ichoice}
\nonumber
\end{eqnarray}
with all others being zero. 

We now show  that with this choice we cancel the non-abelian anomalies.
They cancel if the following condition is satisfied:
\begin{eqnarray}
\sum_{B} \left( I_{A B} +  I_{A  B^*} \right) N_{B}
=0~~;~~
A =a,b
\label{sum=0}
\end{eqnarray}
This follows from Eq. (\ref{nonabe1}) together with Eq. (\ref{aO6}). 
For $A =a$ we get:
\begin{eqnarray}
&&\sum_{B} \left( I_{a B} +  I_{a B^*} \right)N_{B}=
\nonumber \\ 
&& = \left( I_{ab}  + I_{a b^*} \right) N_{b} +   \left( I_{ac}  + I_{a c^*} \right) N_{c}=
\nonumber \\
&& =
2( 1 + 2)  -3 -3 =0
\label{zero62}
\end{eqnarray}
while for $A =b$ we get (remember that $I_{ba} = - I_{ab}$ and $I_{b a^*}=
I_{a b^*}$):
\begin{eqnarray}
\sum_{\beta} \left( I_{b \beta} +  I_{b \beta^*} \right)N_{\beta}=
 \left( I_{ba}  + I_{b a^*} \right) N_{a} +    I_{bd }  N_{d} = 
3( -1  +2)  -3 =0 
\label{zero63}
\end{eqnarray}
Note that, if we had chosed $I_{ab} =3$ and $I_{a b^*} =0$ we would not
 have satisfied the previous anomaly cancellation equation.  This means that,
from the point of view of the anomaly cancellation, the representations $2$ and 
${\bar{2}}$ are not equivalent. 
%The previous numbers insure that there is 
%no non-abelian anomaly
%{$\Longrightarrow$} Tadpole cancellation conditions.
%Find a configuration without tadpoles  \alert{$\Longrightarrow$}
% absence of non-abelian anomalies. 
Furthermore,
the anomaly cancellation requires that the number of generations be equal
to the number of colors. In conclusion, with the choice in Eqs. (\ref{Ichoice}) 
we do not have any non-abelian anomaly. 

We have to check now what happens for the mixed 
$U(1)_{A}- SU(N_{B})_{B}$ 
 anomalies. The coefficient of these anomalies is equal to 
 (see Eq. (\ref{mia1})):
\begin{eqnarray}
{\cal{A}}_{A B} \equiv  \frac{1}{2}
%\sum_{B} 
N_{A}  \left( I_{A  B} +  I_{A B^*} \right)  
\label{mi45}
\end{eqnarray}
We have to compute this quantity for $B=a,b$. For $ B=a$ we get:
\begin{eqnarray}
{\cal{A}}_{ba} =1~~;~~{\cal{A}}_{ca} =0~~;~~{\cal{A}}_{da} =0
\label{betaa}
\end{eqnarray}
and for $B= b$ we get:
\begin{eqnarray}
{\cal{A}}_{ab} =\frac{9}{2}~~;~~{\cal{A}}_{cb} =0~~;~~{\cal{A}}_{db} =\frac{3}{2}
\label{betab}
\end{eqnarray}
They are the coefficients of the anomaly of the various $U(1)$ currents 
involving the second Chern class of both the non-abelian $SU(3)$ and $SU(2)$.
For instance, the $U(1)$ current corresponding to $Q_a$ has the $SU(2)$ 
anomaly, but not the $SU(3)$ anomaly, while that corresponding to 
$Q_b$ has $SU(3)$ anomaly, but not $SU(2)$ anomaly.
{From} the previous coefficients we can read that the $U(1)$'s 
with generators 
\begin{eqnarray}
Q_c ~~~~~~  and ~~~~~Q_a - 3 Q_d  
\label{QQafree}
\end{eqnarray}
are  anomaly free, while
the orthogonal combinations:
\begin{eqnarray}
{\hat{Q}} \equiv 3 Q_a + Q_d~~~;~~~Q_b
\label{QQ}
\end{eqnarray}
are anomalous. In terms of the two non-anomalous $U(1)$'s we can 
construct the generator of the $U(1)$ hypercharge:
\begin{eqnarray}
Q_{Y} = \frac{1}{6} Q_a - \frac{1}{2} Q_c - \frac{1}{2} Q_d
\label{QY}
\end{eqnarray}
that must be anomaly-free. The three orthogonal combinations corresponding to
\begin{eqnarray}
&& Q_b ~~~;~~~ {\hat{Q}}=3 Q_a + Q_d \nonumber \\
&&{\tilde{Q}} = Q_a + \frac{10}{3} Q_c - 3 Q_d
\label{Qano}
\end{eqnarray}
are instead anomalous. Remember that $Q_a$ corresponds to the baryon number
$B$, while $Q_d$ corresponds to the lepton number $L$:
\begin{eqnarray}
Q_a = 3B~~~;~~~{\hat{Q}}=3 Q_a + Q_d = 3 (B-L) 
\label{Qano2}
\end{eqnarray}
$Q_b$ corresponds instead to a Peccei-Quinn symmetry that has a $SU(3)$ mixed anomaly as follows from the fact that ${\cal{A}}_{ba}$ in Eq. (\ref{betaa}) 
is different from zero. However, at this point we have the problem that those $U(1)$
correspond to  anomalous gauge symmetries . Furthermore, $Q_a$ and ${\hat{Q}}$
are in our case gauge symmetries and not global symmetries as in the SM. 
In the following we will show how these two problems are solved. 

We have seen that the $U(1)$ hypercharge in Eq. (\ref{QY}) is anomaly free, 
while the other three $U(1)$ are anomalous.  In string theory the anomaly comes 
 from the one-loop planar diagram that, in the field theory limit, 
reduces to the well known triangular diagram of anomalies. However, in string
theory we have also a non-planar diagram that actually cancels the anomaly
of the planar one~\cite{9808139}. In particular, this cancellation 
comes from a term of 
the non-planar diagram that corresponds to the exchange, in the closed 
string channel, of a RR $C_2$ field that is coupled, on the one side, to the 
gauge field of anomalous $U(1)$ and on the other side to the two gauge fields
of the non-abelian group. One of the couplings  is divergent in the limit
$\alpha' \rightarrow 0$, while the other goes to zero in such a way that their product
is independent of $\alpha'$, giving a contribution that exactly cancels that of
the planar diagram. This is called Green-Schwarz mechanism because it is
the same that eliminates the gauge anomaly in ten-dimensional type I
string theory. In conclusion, also the anomalies of the other three $U(1)$
cancel if one takes into account the contribution to the anomaly of the
non-planar diagram.  This is a pure stringy effect although it gives a 
contribution that is not vanishing in the field theory limit.

If this were the end of the story, then we will be left with three additional
gauge $U(1)$'s and not with just one as in the SM.

However, this is not  true because string theory contains an additional
mechanism, discovered in string theory by Cremmer and Scherk~\cite{CS}, 
that is called
St{\"{u}}ckelberg mechanism, according to which the three extra $U(1)$ get
a non-zero mass.   This is again due to the coupling of the $U(1)$ gauge fields
with a RR field $C_2$ that diverges as $\frac{1}{\sqrt{\alpha'}}$ that, together
with the kinetic terms for the gauge field and for $C_2$, provides a non-zero mass
to the $U(1)$ gauge field. In conclusion, the three extra $U(1)$ get a
non-zero mass of the order   $\frac{1}{\sqrt{\alpha'}}$ and this implies that
the original three local $U(1)$ become three global ones. This is the origin
of the global symmetries, $B-L$ and baryon number, of the SM 
in this string extension of the SM.  Finally, we must be careful that the
gauge boson of the hypercharge $U(1)$ does not get a mass. This can be
done imposing an extra condition that can be found in Ref.~\cite{0307262}
together with a more complete description of the model. In this way one
obtains a string extension of the SM with only one additional 
particle, the right-handed neutrino.  

The three   $U(1)$'s  whose gauge boson got a mass, are exact 
global symmetries at each order of  string perturbation theory.
Therefore the baryon and lepton numbers are exactly preserved and
Majorana neutrino masses are  not allowed at each order 
of perturbation theory. These symmetries, however,  can be broken by 
instantons and this has been proposed as a way to give a Majorana mass
to the neutrinos~\cite{0609191,0609213}. 
It would be very interesting if this effect could be  due to
 pure stringy effects that disappear in the 
field theory limit ($\alpha' \rightarrow 0$)!

\vspace{.5cm}
%\noindent
I thank R. Marotta for many discussions and collaboration on various 
issues discussed in this paper and for a critical reading of the manuscript.
I thank also F. Marchesano for some email exchanges.

\end{document}